\makeatletter
\input{filecontents.sty}
\makeatother

\documentclass[%
titlepage,
titlepageabstract,
secnumarabic,
showpacs,
twocolumn,
twoside,
floatfix,
a4paper,
prl]{oxprepr}

\def\Preprint{}
\makeatletter
\ifx\@undefined\Preprint
  \def\JournalPreprint#1#2{#1}
\else
  \def\JournalPreprint#1#2{#2}
\fi
\makeatother

\JournalPreprint{\documentclass[twocolumn,prl,showpacs,floatfix]{revtex4}}{}

\usepackage[american]{babel}
\usepackage{amsfonts}
\usepackage{amsmath}
\usepackage{amssymb}
\usepackage{graphicx}
\usepackage{color}
\usepackage[pdftitle={Topological phase transition in complex networks},
pdfauthor={Heiko Bauke and David Sherrington},
pdfcreator={},
pdfsubject={complex networks, non-equilibrium phase transition},
pdfproducer={},
pdfkeywords={complex networks, non-equilibrium phase transition},
hyperfootnotes=false,naturalnames]{hyperref}
\graphicspath{{figures/}}

\IfFileExists{date.tex}{%

}

\newcommand*{\D}{\mathrm{d}}
\newcommand*{\bigO}[1]{\mathcal{O}\left(#1\right)}

\begin{document}

\title{Topological phase transition in complex networks}

\author{Heiko Bauke}
\email[E-mail:]{heiko.bauke@physics.ox.ac.uk}
\author{David Sherrington}
\affiliation{Rudolf Peierls Centre for Theoretical Physics, University of
  Oxford, 1~Keble~Road, Oxford, OX1~3NP, United Kingdom}

\date{\today}

\begin{abstract}
  Preferential attachment is a central paradigm in the theory of complex
  networks.  In this contribution we consider various generalizations of
  preferential attachment including for example node removal and edge
  rewiring. We demonstrate that generalized preferential attachment networks
  can undergo a topological phase transition. This transition separates
  networks having a power-law tail degree distribution from those with an
  exponential tail. The appearance of the phase transition is closely related
  to the breakdown of the continuous variable description of the network
  dynamics.
\end{abstract}

\pacs{89.75.Fb, 89.75.Da}
% 89.75.Da 	Systems obeying scaling laws
% 89.75.Fb 	Structures and organization in complex systems

\maketitle

\noindent%
Complex networks are found in nature, socio-economic systems, technical
infrastructures, and countless other fields. The macroscopic properties of
such networks emerge from the microscopic interaction of many individual
constituents. Various models of complex networks have been proposed
\cite{Dorogovtsev:Mendes:2003:Evolution_of_networks,Newman:2004:complex_networks,Boccaletti:etal:2006,Newman:Barabasi:Watts:2006:Networks}
and statistical mechanics of complex networks has become an established branch
of statistical physics. Since complex networks are non-equilibrium systems
they do not have to obey detailed balance and may show many fascinating
features not found in equilibrium systems.

In order to explain the power-law degree distribution of many complex networks
Barab{\'a}si and Albert \cite{Barabasi:Albert:1999:Scaling_in_Random_Networks}
introduced a model that combines network growth and preferential
attachment. Preferential attachment means that new nodes enter a network by
attaching preferably to nodes having a high degree. The probability that a new
node $u$ establishes an edge to an old node $v$ with degree $k_v$ is
determined by a node attractiveness function $A_k$; it is given by
$p_\mathrm{add}(v)=A_{k_v}/\sum_w A_{k_w}$.
% \begin{equation}
%   \label{eq:p_add}
%   p_\mathrm{add}(v)=\frac{A_{k_v}}{\sum_w A_{k_w}} \,.
% \end{equation}
In the original Barab{\'a}si-Albert model
\cite{Barabasi:Albert:1999:Scaling_in_Random_Networks} the attractiveness of a
node is strictly-proportional to its degree, viz.\ $A_k = k$.

The linear preferential attachment function of the Barab{\'a}si-Albert model was
introduced as an \emph{ad hoc} ansatz without any fundamental justification
and many generalizations are conceivable, including attractiveness functions
with small perturbations,
\begin{equation}
  \label{eq:A_k_eps}
  A_k = k + \varepsilon(k, k^*) 
  \quad\text{with}\quad
  \varepsilon(k, k^*)\in\bigO{1}\,,
\end{equation}
random node removal, rewiring, etc. Here we will consider generalized
preferential attachment models that include tunable parameters that allow to
adjust the relative importance of node removal, rewiring, etc. These
parameters define the phase space of a model.

We will show that generalizations of the Barab{\'a}si-Albert model as considered
here can dramatically affect the degree distribution; it falls into one of two
possible universality classes. At each point of the phase space the
universality class is uniquely determined by a parameter $\beta$. For
$\beta>0$ the tail of the degree distribution is given by a power-law, while
if $\beta<0$ it is an exponential. The two universality classes are separated
by a topological phase transition. The behavior at the phase transition
($\beta=0$) depends on the specific model.

We claim that our results are valid for a large class of generalized
preferential attachment models, but let us consider two specific examples for
illustrative purposes.
\begin{description}
\item[Model I] At each time step with probability~$r$ a randomly chosen node
  is removed from the network and with probability~$1$ a new node is added to
  the network by establishing $m$ links to $m$ old nodes. These nodes are
  chosen randomly by preferential attachment with the attractiveness function
  (\ref{eq:A_k_eps}). The perturbation function may be any bounded function
  $\varepsilon(k, k^*)\in\bigO{1}$ such that $A_k\ge0$, e.\,g.
  \begin{equation}
    \label{eq:eps1}
    \varepsilon(k, k^*) =
    \begin{cases}
      k^* k/m & \text{if $k\le m$} \\
      k^* & \text{else.}
    \end{cases}
  \end{equation}
  For $k\ge m$ function (\ref{eq:eps1}) implements a constant shift of size
  $k^*$.  The network starts to grow at time $t=0$ with $N_0$ nodes. We are
  interested in behaviour where the number of nodes is much larger than
  unity. For $r=1$ the number of nodes is conserved so $N_0\gg1$, but for
  $r<1$ the average number of nodes grows with time and one may start with
  $N_0 = \bigO{1}$.
\item[Model II] We start from a small initially given network. At each time
  step one of three different operations is performed with probability $p$,
  $q$, or $r$, respectively, such that $p+q+r=1$ and
  $0<p,q,r<1$.\footnote{This model was considered earlier in
    \cite{Albert:Barabasi:2000:Universality}, but the analysis of this model
    was incorrect. In the rewiring step it was not taken into account that if
    one follows the edge from $u$ to $v$ the probability that $v$ has degree
    $k_v$ is proportional to $k_v$.}
  \begin{description}
  \item[Edge insertion (probability $p$)] There are $m$ new edges
    inserted. For each edge one end $u$ is chosen randomly from the set of
    nodes without any preference, while the other end $v$ is selected
    preferentially with probability proportional to \mbox{$A_{k_v}=k_v+1$}.
  \item[Rewiring (probability $q$)] $m$ links are rewired by selecting a node
    $u$ (with non-zero degree) and one of its neighbors $v$. The edge between
    $u$ and $v$ is removed and a new edge from $u$ to $v'$ is added, where
    $v'$ is chosen with probability proportional to $A_{k_{v'}}=k_{v'}+1$.
  \item[Node addition (probability $r$)] A new node is added to the network by
    establishing $m$ links to $m$ old nodes $v$ which are chosen with
    probability proportional to $A_{k_{v}}=k_{v}+1$.
  \end{description}
\end{description}

\begin{figure}
  \centering%
  \includegraphics[scale=0.42]{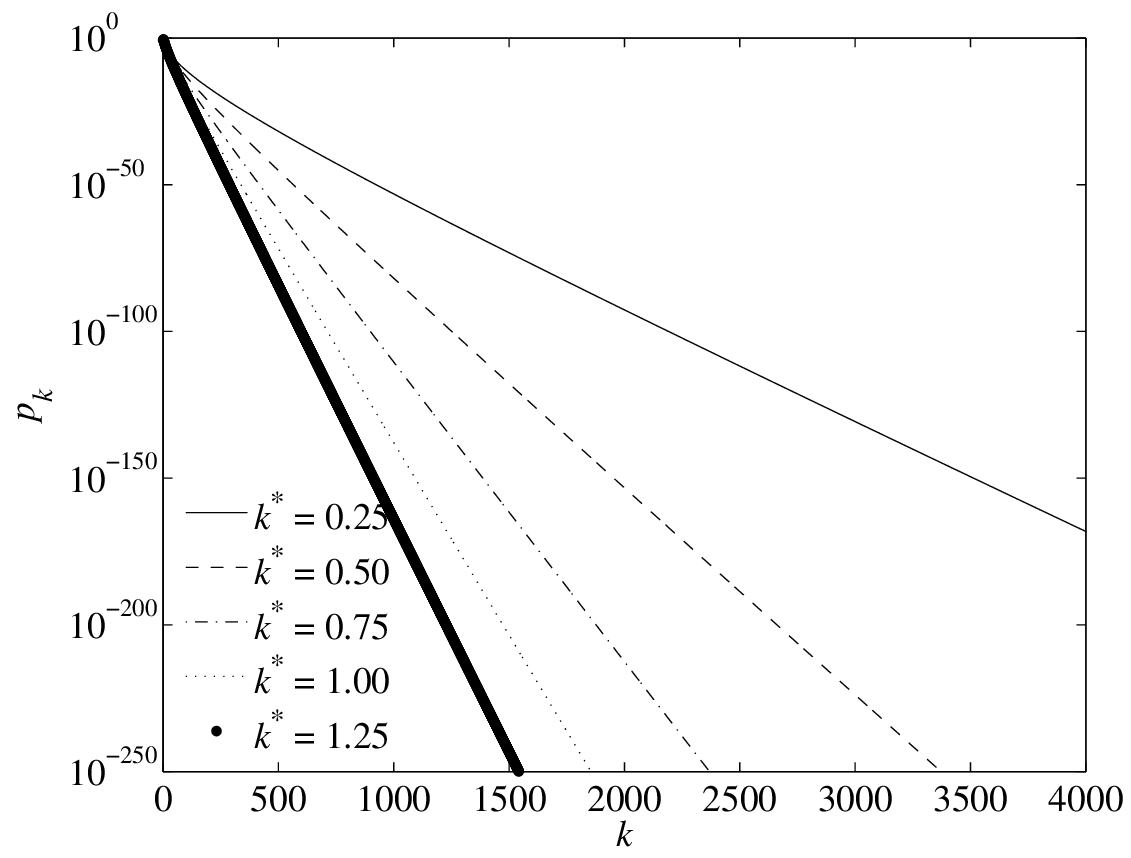}\\
  \raisebox{60mm}[0mm][0mm]{(a)}\hfill\null\\%
  \includegraphics[scale=0.42]{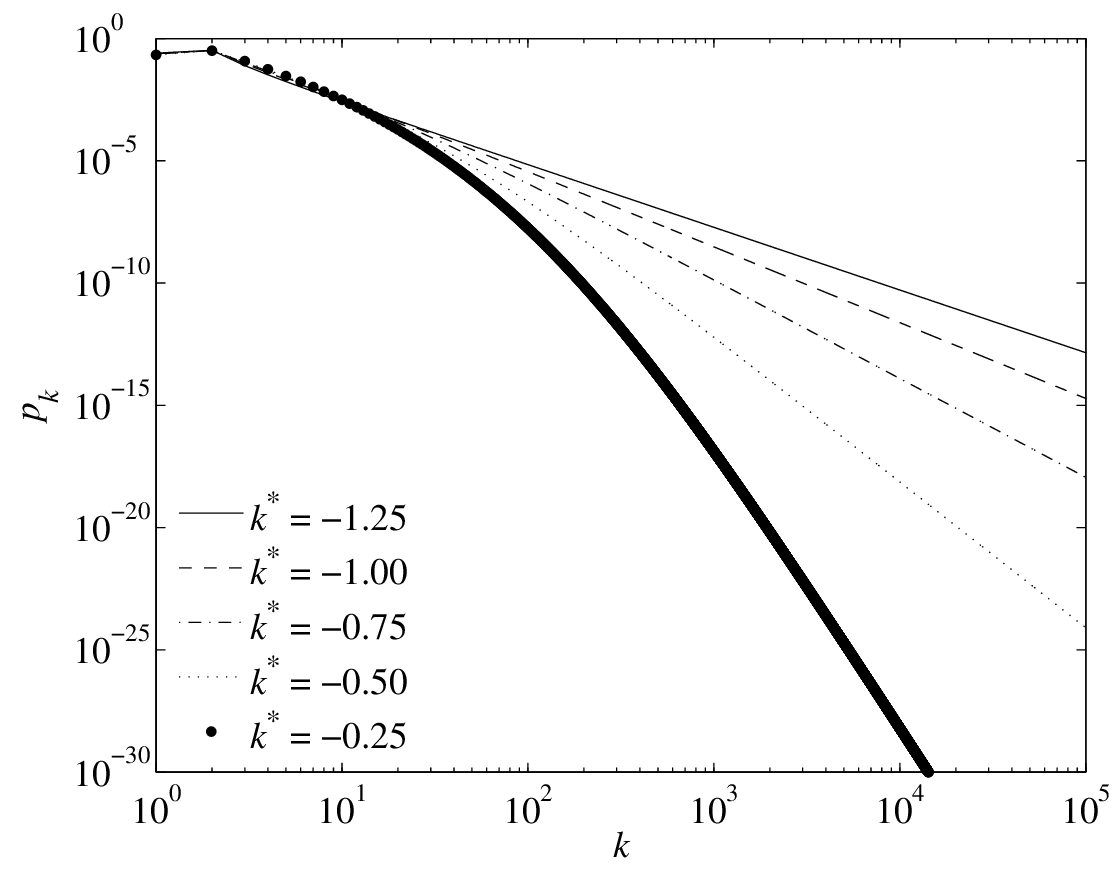}\\
  \raisebox{60mm}[0mm][0mm]{(b)}\hfill\null\vspace{-2ex minus 2ex}%
  \caption{Degree distribution for model~I with the perturbation function
    (\ref{eq:eps1}), node deletion rate $r=1$, and $m=2$; sub-figure (a)
    exponential case, sub-figure (b) power-law case.}
  \label{fig:p1_k}
\end{figure}
\noindent Let $p_k$ denote the degree distribution of a generalized
preferential attachment model, i.\,e., the relative number of nodes having
degree $k$ averaged over many network instances. This distribution can be
calculated by solving a master equation with the boundary values $p_{-1}=0$
and $p_{k_\mathrm{max}+1} = 0$, where $k_\mathrm{max}$ denotes the maximal
possible degree in the network. The thermodynamic limit corresponds to
$k_\mathrm{max}\to\infty$.  For model~I the master equation reads
\cite{Moore:Ghoshal:Newman:2006:Exact_solutions}
\begin{equation}
  \label{eq:masterI}
  \delta_{k,m} =
  - m \frac{A_{k-1}}{\bar A} {p_{k-1}}
  + \left(m\frac{A_k}{\bar A} + rk + 1\right){p_k}
  - r(k+1){p_{k+1}}\,,
\end{equation}
which contains the mean attractiveness 
\begin{equation}
  \label{eq:norm}
  \bar A = \sum_k A_k p_k \,.
\end{equation}
The simultaneous numerical solution of equations (\ref{eq:masterI}) and
(\ref{eq:norm}) gives either a degree distribution with a power-law tail or an
exponential tail, as shown in \figurename~\ref{fig:p1_k}. For model~II we can
write down a similar master equation and its solution obeys either a power-law
or an exponential tail, too, depending on the parameters $p$, $q$, $r$, and
$m$.

Analysis of generalized preferential network models in terms of a continuous
variable approximation \cite{Dorogovtsev:Mendes:2001:Scaling_properties} sheds
light on the appearance of the phase transition. We will outline the generic
mechanism that leads to the topological phase transition
% from power-law to exponential degree distributions
first and we will consider specific models later on.

In a continuous variable approximation degrees are treated as continuous
variables and random quantities are represented by their means. The degree
distribution $p_k$ is given by
\cite{Sarshar:Roychowdhury:2004:complex_ad_hoc_networks}
\begin{subequations}
  \label{eq:p_k}
\begin{equation}
  \label{eq:p_ka}
  p_k = \frac{D(i_k, t)}{N(t)} \left|
    \frac{\partial k(i, t)}{\partial i}
  \right|_{i=i_k}^{-1}\,.
\end{equation}
\end{subequations}
Here $N(t)$ denotes the size of the network at time $t$. Node addition and
removal lead to an effective network model-de\-pend\-ent growth rate $\alpha$,
i.\,e., $N(t)=\alpha t+N_0$. The quantity $D(i,t)$ takes into account random
node removal at rate $\rho$. It is the probability that a node that entered
the network at time $i$ is still present at time $t\ge i$ and is given by the
initial condition $D(i, i)=1$ and the equation
\begin{subequations}
  \label{eq:D_t}
\begin{equation}
  \label{eq:D_ta}
  \frac{\partial D(i, t)}{\partial t} = -\rho \frac{D(i, t)}{N(t)}\,.
\end{equation}
\end{subequations}
The function $k(i, t)$ denotes the mean degree 
% (ensemble average over different network realizations)
of a node at time $t\ge i$ that entered the network at time $i$ and has not
yet disappeared from the network. The time $i_k\le t$ is given by the solution
of the equation $k(i_k, t)=k$. The quantity $\left|{\partial k(i,
    t)}/{\partial t}\right|_{i=i_k}^{-1}$ can be interpreted as the number of
nodes with degree $k$ at time $t$ (neglecting node deletion).

For generalized preferential attachment networks the evolution of $k(i, t)$ is
given by an equation which has the generic form
\begin{subequations}
  \label{eq:k_d_t}
  \begin{equation}
  \label{eq:k_d_ta}
    \frac{\partial k(i, t)}{\partial t} = 
    \beta\frac{k(i, t)}{N(t)} + \frac{f[k(i, t)]}{N(t)}\,.
  \end{equation}
\end{subequations}
The term $\beta{k(i, t)}/{N(t)}$ is the hallmark of preferential attachment
\cite{Barabasi:Albert:1999:Scaling_in_Random_Networks}, while ${f[k(i,
  t)]}/{N(t)}$ with $f(x)\in\bigO{1}$ appears as a consequence of
perturbations of the strictly linear attractiveness function, edge rewiring,
node deletion, or other processes.

Before analyzing the equations (\ref{eq:p_ka}), (\ref{eq:D_ta}), and
(\ref{eq:k_d_ta}) in detail it is useful to introduce a new time scale
$t'$. For $\alpha>0$ we set $t=\exp(\alpha t')$ and in the case $\alpha=0$ we
set $t=t'N_0$. With this new time scale the number of node additions per
unit-time is proportional to the network size. Now equations (\ref{eq:p_ka}),
(\ref{eq:D_ta}), and (\ref{eq:k_d_ta}) read (omitting primes)
\begin{gather}
  \tag{\ref{eq:p_k}b}\label{eq:p_kb}
  p_k = \frac{D(i_k, t)}{\exp[\alpha(t-i_k)]} \left|
    \frac{\partial k(i, t)}{\partial i}
  \right|_{i=i_k}^{-1}\,,\\
  \tag{\ref{eq:D_t}b}\label{eq:D_tb}
  \frac{\partial D(i, t)}{\partial t} = -\rho D(i, t)\,,\\
  \tag{\ref{eq:k_d_t}b}\label{eq:k_d_tb}
  \frac{\partial k(i, t)}{\partial t} = 
  \underbrace{\beta k(i, t)}_
  {{\mbox{I}}} + 
  \underbrace{f[k(i, t)]}_
  {{\mbox{II}}}\,.
\end{gather}
The explicit dependence on the network size has dropped out of equations
(\ref{eq:p_kb}), (\ref{eq:D_tb}), and (\ref{eq:k_d_tb}) and we need not to
distinguish between growing ($\alpha>0$) and non-growing ($\alpha=0$)
networks.

The occurrence of the phase transition can be pinned down to a change in the
solution structure of (\ref{eq:k_d_tb}). We have to distinguish the three
cases $\beta>0$, $\beta=0$, and $\beta<0$. First let us consider
$\beta>0$. Because $f[k(i, t)]\in\bigO{1}$ equation (\ref{eq:k_d_tb}) is
dominated by term~I.  For $k(i, t)\gg1$ the leading term of the solution of
(\ref{eq:k_d_tb}) is given by
\begin{equation}
  \label{eq:k_i_t}
  k(i, t) \sim \exp\left[\beta(t-i)\right]
\end{equation}
and there follows
\begin{equation}
  \label{eq:i_k}
  i_k = t - \frac{\ln k}{\beta} \,.
\end{equation}
Plugging (\ref{eq:k_i_t}), (\ref{eq:i_k}), and the solution of
(\ref{eq:D_tb}),
\begin{equation}
  \label{eq:D_i_t}
  D(i, t) = \exp[-\rho(t-i)]\,,
\end{equation}
into (\ref{eq:p_kb}) we find that the combination of exponentials with
different exponents \cite{Newman:2005:power-laws} yields a power-law degree
distribution $p_k\sim k^{-\gamma}$ with exponent
\begin{equation}
  \label{eq:gamma}
  \gamma = \frac{\rho+\alpha}{\beta} + 1\,.
\end{equation}
The power-law exponent (\ref{eq:gamma}) is determined entirely by three rates,
the network growth rate $\alpha$, the node degree growth rate $\beta$, and the
node deletion rate $\rho$. It diverges as $\beta\to0$.

It turns out that the exponential growth of a node's degree (\ref{eq:k_i_t})
is the key ingredient for the emergence of a power-law degree
distribution. For $\beta=0$ this ingredient is lost and the phase transition
appears. The equation $\beta=0$ defines the critical phase line that separates
the two universality classes. At the phase transition term~I in equation
(\ref{eq:k_d_tb}) vanishes %,
% \begin{equation}
%   \frac{\partial k(i, t)}{\partial t} = f[k(i, t)] \,,
% \end{equation}
and as a consequence the qualitative behavior of the solution of
(\ref{eq:k_d_tb}) is determined by term~II and not universal.  Because
$f(x)\in\bigO{1}$ function $k(i, t)$ must grow sub-exponentially. The solution
of (\ref{eq:k_d_tb}) is given by the implicit equation
\begin{equation}
  \label{eq:k_i_t_1}
  \int_m^{k(i, t)} \frac{1}{f(k')}\,\D k' = t-i\,.
\end{equation}

For $\beta<0$ term~I dominates equation (\ref{eq:k_d_tb}) again. But now the
continuous variable approximation breaks down. In fact, the time (\ref{eq:i_k})
becomes greater than $t$; that means it is in the future. The  continuous
variable approximation is not able to predict the correct degree distribution
for $\beta<0$.%\enlargethispage{\baselineskip}

Let us see how our generic continuous variable analysis translates to specific
models. In model~I in each time step a node is added to the network and with
probability $\rho=r$ a node is removed from the network, as a consequence its
size grows at rate $\alpha=(1-r)$. The function $k(i, t)$ is given by
\begin{equation}
  \label{eq:k_d_t_I}
  \frac{\partial k(i, t)}{\partial t} = 
  m\frac{k(i, t) + \varepsilon[k(i, t), k^*]}{\bar A N(t)} - 
  r\frac{k(i, t)}{N(t)}\,.
\end{equation}
\begin{figure}
  \centering
  \includegraphics[scale=0.42]{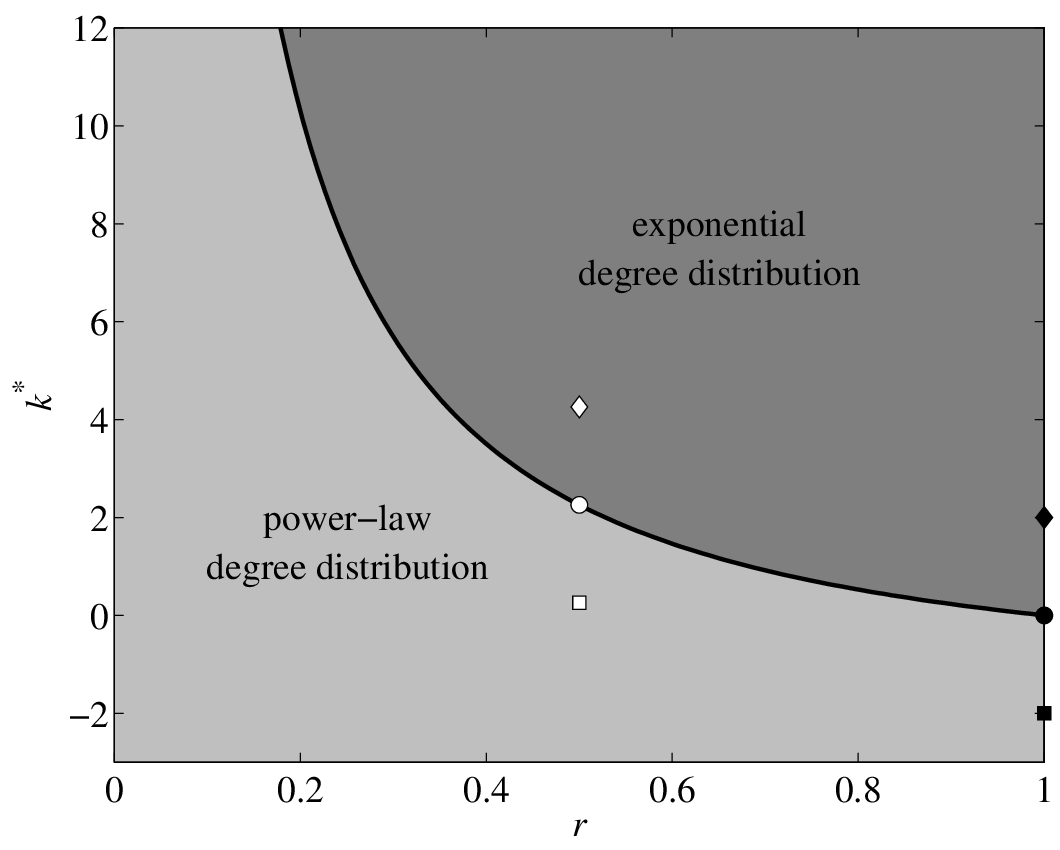}%
  \caption{Phase diagram of the degree distribution of model~I with the
    perturbation function (\ref{eq:eps1}) and $m=3$. Symbols indicate
    locations in the phase space for data shown in
    \figurename~\ref{fig:node_evolution}.}
  \label{fig:phase_diagram_I}
\end{figure}
\begin{figure}[b]
  \centering%
  \includegraphics[scale=0.42]{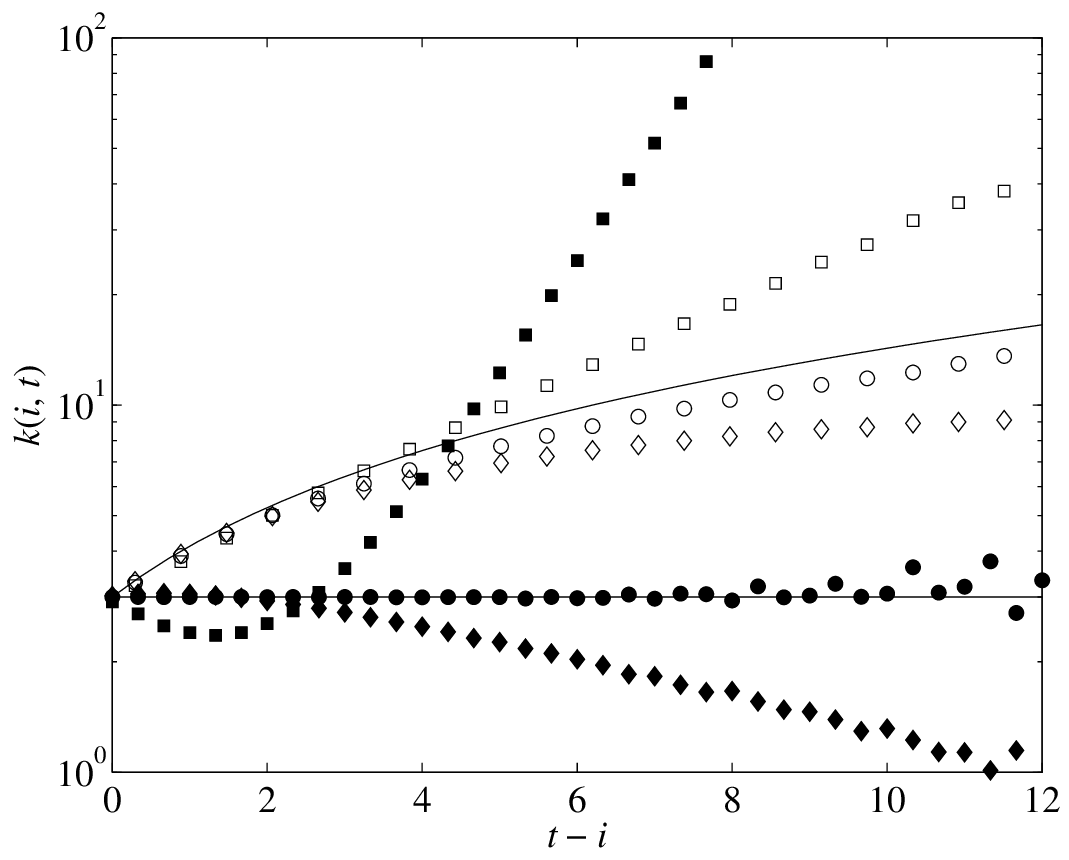}
  \caption{Mean degree of a node at time $t\ge i$ (at rescaled time scale) for
    model~I networks with perturbation function (\ref{eq:eps1}), $m=3$, and
    removal rates $r=0.5$ (open symbols) and $r=1$ (solid symbols),
    respectively. The diagram shows networks at the phase transition, in the
    exponential regime, as well as in the power-law regime; see
    \figurename~\ref{fig:phase_diagram_I} for the corresponding position in
    the phase diagram. Symbols show binned data obtained by a direct
    simulation of the network dynamics, lines are given by analytic
    computations of $k(i,t)$ at the phase transition (\ref{eq:k_i_t_1}).}
  \label{fig:node_evolution}
\end{figure}
The first term takes into account node addition via preferential attachment
with attractiveness function (\ref{eq:A_k_eps}) and the second term
corresponds to random node deletion. After reordering equation
(\ref{eq:k_d_t_I}) and comparing with (\ref{eq:k_d_ta}) we find
$\beta=[m/(r\bar A) - 1]r$. For $\beta>0$ the degree distribution is given by
a power-law with exponent $\gamma=1/\{[m/(r\bar A) - 1]r\}+1$. In order to
determine the universality class of the degree distribution we have to
calculate $\beta$ and $\bar A$ by solving equations (\ref{eq:masterI}) and
(\ref{eq:norm}) numerically. If we choose the perturbation function
(\ref{eq:eps1}) then the phase space is spaned by $r$ and $k^*$, where the
critical phase line is given by $[m/(r\bar A) - 1]r=0$, see
\figurename~\ref{fig:phase_diagram_I}. Direct simulations of the network
dynamics show that in the universality class of power-law degree distributions
$k(i, t)$ grows indeed exponentially, as predicted by (\ref{eq:k_i_t}), while
in the exponential class $k(i, t)$ grows sub-ex\-po\-nent\-ially, see
\figurename~\ref{fig:node_evolution}.

Moreover, our findings for model~I put some known results for specific
variations of the Barab{\'a}si-Albert model into a unifying framework.  For a
vanishing node deletion rate $r=0$ and \mbox{$\varepsilon(k, k^*)=k^*$} the
mean attractiveness equals $\bar A = 2m+k^*$ and for any $k^*$ the degree
distribution has a power-law tail with $\gamma=3+k^*/m$
\cite{Dorogovtsev:Mendes:Samukhin:2000:Structure_of_Growing_Networks}.  In the
special case $\varepsilon(k, k^*)=0$ with node deletion rate $0<r<1$ the mean
attractiveness $\bar A$ equals the mean degree of the network, which is given
by \mbox{$2m/(1+r)$}. It follows that the degree distribution is in the
power-law class with $\gamma=\mbox{$(3-r)$}/\mbox{$(1-r)$}$
\cite{Sarshar:Roychowdhury:2004:complex_ad_hoc_networks,Moore:Ghoshal:Newman:2006:Exact_solutions}.

\begin{figure}
  \centering
  \includegraphics[scale=0.42]{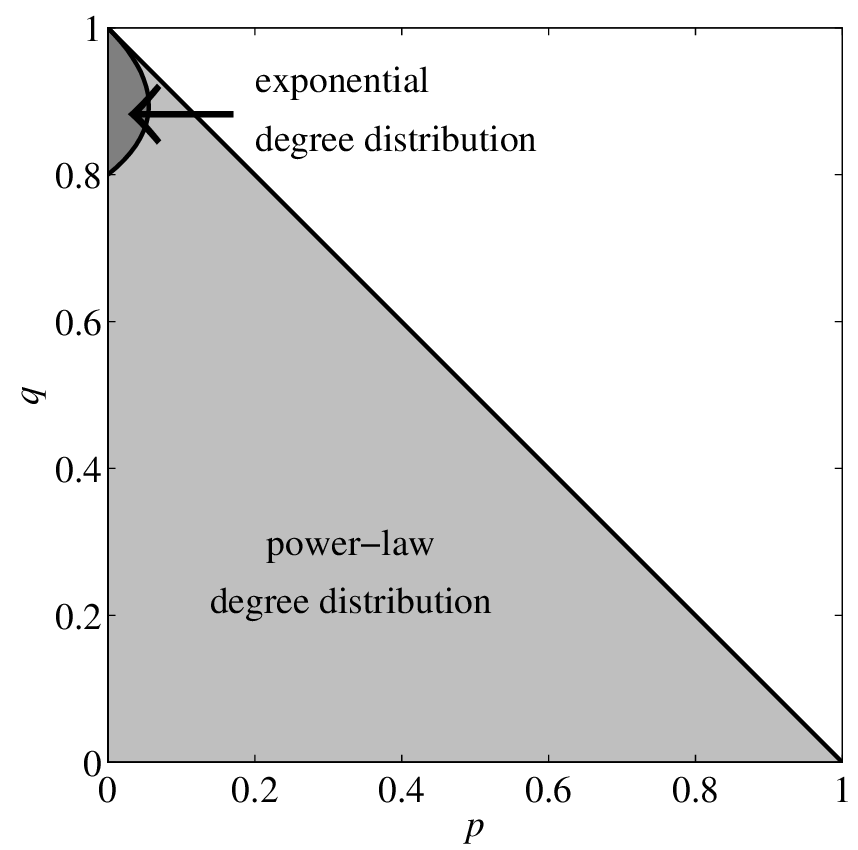}%
  \caption{Phase diagram of model~II for $m=2$.}
  \label{fig:phase_diagram_II}
\end{figure}

Model~II can be analyzed in a similar way. In this model the network size
grows at rate $\alpha=r=1-p-q$, and the evolution equation for $k(i, t)$ is
given by
\begin{multline}
  \label{eq:k_d_t_II}
  \frac{\partial k(i, t)}{\partial t} = 
  \frac{mp}{N(t)}+\frac{mp[k(i, t)+1]}{N(t)(\bar k +1)} \\
  \null- \frac{mqk(i, t)}{N(t)\bar k} + \frac{mq[k(i, t)+1]}{N(t)(\bar k +1)} +
  \frac{mr[k(i, t)+1]}{N(t)(\bar k +1)}\,,
\end{multline}
where $\bar k=2m(1-q)/r$ denotes the mean degree of the network. After some
rearranging of equation (\ref{eq:k_d_t_II}) and comparing with
(\ref{eq:k_d_ta}) we identify
\begin{equation*}
  \beta = \frac{1}{2} \frac{(1-p-q)[-(2m+1)q^2 + (4m+1-p)q - 2m]}
  {(1-q)[(2m+1)q-(2m+1)+p]}\,.
\end{equation*}
So, for $\beta>0$ the degree distribution will follow a power-law
with exponent
\begin{equation*}
  \gamma = 2\frac{(1-q)[(2m+1)q-(2m+1)+p]}{-(2m+1)q^2 + (4m+1-p)q - 2m} + 1
\end{equation*}
and the critical phase line is determined by the equation \mbox{$(2m+1)q^2 -
  (4m+1-p)q + 2m=0$}, as shown in \figurename~\ref{fig:phase_diagram_II}.

To summarize, we have investigated a broad class of generalized preferential
attachment models and shown that small perturbations of the attractiveness
function (\ref{eq:A_k_eps}), node removal, or edge rewiring cannot only change
the power-law exponent of the original Barab{\'a}si-Albert model
\cite{Barabasi:Albert:1999:Scaling_in_Random_Networks}. These variations can
even lead to a qualitative change of the degree distribution from a power-law
to an exponential.

It had been argued \cite{Moore:Ghoshal:Newman:2006:Exact_solutions} ``that net
growth is one of the fundamental requirements for the generation of power-law
degree distributions'' of preferential attachment models. However, we have
shown, that it is not the growth of the network itself that leads to a
power-law degree distribution. The mean degree of individual nodes has to grow
sufficiently strongly, i.\,e.\ $\beta>0$, to generate a power-law degree
distribution.

We also demonstrated that negative perturbations of the linear attractiveness
function as in model~I have a tendency to stabilize power-laws. In the light
of this fact and the ubiquity of power-laws in nature it would be interesting
to know if in real world networks such negative perturbations do exist.

\vspace*{\fill}%
\acknowledgments This work has been sponsored by the
European Community's FP6 Information Society Technologies programme under
contract IST-001935, EVERGROW.

\bibliography{phase_transition}

\end{document}